\documentclass[twocolumn]{aastex62}

\usepackage[inline]{enumitem}
\newcommand{\deltanu}{\mbox{$\langle \Delta\nu \rangle$}}

\shorttitle{The red-giant branch bump revisited}
\shortauthors{Khan et al.}

\begin{document}

\title{The Red-Giant Branch Bump Revisited:\\ Constraints on Envelope Overshooting in a Wide Range of Masses and Metallicities}

\correspondingauthor{Saniya Khan}
\email{sxk1008@bham.ac.uk}

\author{Saniya Khan}
\affiliation{School of Physics and Astronomy, University of Birmingham, Edgbaston, Birmingham, B15 2TT, UK}
\affiliation{Stellar Astrophysics Centre, Department of Physics and Astronomy, Aarhus University, Ny Munkegade 120, 8000 Aarhus C, Denmark}

\author{Oliver J. Hall}
\affiliation{School of Physics and Astronomy, University of Birmingham, Edgbaston, Birmingham, B15 2TT, UK}
\affiliation{Stellar Astrophysics Centre, Department of Physics and Astronomy, Aarhus University, Ny Munkegade 120, 8000 Aarhus C, Denmark}

\author{Andrea Miglio}
\affiliation{School of Physics and Astronomy, University of Birmingham, Edgbaston, Birmingham, B15 2TT, UK}
\affiliation{Stellar Astrophysics Centre, Department of Physics and Astronomy, Aarhus University, Ny Munkegade 120, 8000 Aarhus C, Denmark}

\author{Guy R. Davies}
\affiliation{School of Physics and Astronomy, University of Birmingham, Edgbaston, Birmingham, B15 2TT, UK}
\affiliation{Stellar Astrophysics Centre, Department of Physics and Astronomy, Aarhus University, Ny Munkegade 120, 8000 Aarhus C, Denmark}

\author{Beno\^{i}t Mosser}
\affiliation{LESIA, Observatoire de Paris, PSL Research University, Universit\'e Pierre et Marie Curie, Universit\'e Denis Diderot, 92195 Meudon Cedex, France}

\author{L\'eo Girardi}
\affiliation{INAF - Osservatorio Astronomico di Padova, Vicolo dell'Osservatorio 5, I-35122 Padova, Italy}

\author{Josefina Montalb\'an}
\affiliation{Dipartimento di Fisica e Astronomia Galileo Galilei, Universit\`a di Padova, Vicolo dell'Osservatorio 3, I-35122 Padova, Italy}

\begin{abstract}

The red-giant branch bump provides valuable information for the investigation of the internal structure of low-mass stars. Because current models are unable to 
accurately predict the occurrence and efficiency of mixing processes beyond convective boundaries, one can use the luminosity of the bump --- a diagnostic of the maximum extension of the convective envelope during the first-dredge up --- as a calibrator for such processes. By combining asteroseismic and spectroscopic constraints, we expand the analysis of the bump to masses and metallicities beyond those previously accessible using globular clusters. Our dataset comprises nearly 3000 red-giant stars observed by {\it Kepler} and with APOGEE spectra. Using statistical mixture models, we are able to detect the bump  
in the average seismic parameters $\nu_{\rm max}$ and $\deltanu$, 
and show that its observed position reveals general trends with mass and metallicity in line with expectations from models. 
Moreover, our analysis indicates that standard stellar models underestimate the depth of efficiently mixed envelopes. The inclusion of significant overshooting from the base of the convective envelope, with an efficiency that increases with decreasing metallicity, allows 
to reproduce the observed location of the bump. Interestingly, this trend was also reported in previous studies of globular clusters.
\end{abstract}



\section{Introduction}

The red-giant branch bump (RGBb) is a key observable that allows investigation of the internal structure of low-mass stars. It corresponds to a temporary drop in luminosity as a star evolves on the red-giant branch (RGB), leading to a local maximum in the luminosity function. The occurrence of the bump is related to the hydrogen-burning shell approaching and eventually advancing through the chemical composition gradient left over by the convective envelope at its maximum depth \citep[see, e.g.,][and references therein]{Christensen-Dalsgaard2015}. Since current stellar models are unable to accurately predict the occurrence and efficiency of mixing processes beyond convective boundaries, one can use the luminosity of the RGBb as a calibrator for such processes.

An improved description of mixing beyond convective envelopes has wide-ranging applications, e.g. from predicting the dredge-up efficiency on the RGB to a more accurate calibration of mass (hence age) of RGB stars based on the carbon-to-nitrogen ratio \citep{Salaris2015}. Also, better insights into the physics of mixing processes beyond the Schwarzschild border would have implications for the properties of the tachocline \citep{Christensen-Dalsgaard2011} and the lithium depletion \citep[e.g.][]{Baraffe2017} in Sun-like stars, the evolution of asymptotic-giant branch stars \citep{Herwig2000, Marigo2007}, as well as the onset of blue loops in intermediate and massive stars \citep{Alongi1991, Tang2014}.

The RGBb has been known for a long time: first theoretically, through models of low-mass stars showing a temporary luminosity decrease during the evolution on the RGB \citep{Thomas1967, Iben1968}; then observationally, with its first empirical confirmation in the Galactic Globular Cluster (GGC) 47 Tuc by \citet{King1985}. In particular, the RGBb characteristic luminosity is a diagnostic of the maximum extension of the convective envelope reached during the first dredge-up. Despite the wealth of theoretical and observational investigations, there is an ongoing debate as to a discrepancy between standard models' predictions and observations of the RGBb brightness in GCs \citep{FusiPecci1990, Cassisi1997a, Zoccali1999, Riello2003, Bjork2006}. However, recent studies seem to converge on the significance of this discrepancy and identify overshooting from the bottom of the convective envelope as a plausible solution to reproduce the observational constraints (\citealt{DiCecco2010}; \citealt{Cassisi2011}; \citealt{Troisi2011}; \citealt{Joyce2015} and \citealt{Fu2018}, who used the most recent empirical RGBb magnitudes from \citealt{Nataf2013}). 
Alternative explanations explored in the literature are, for instance, differences in the chemical composition profile and / or opacities leading to a deeper convective region, together with other types of additional mixing \citep[see, e.g.,][]{Bjork2006, Cassisi2011}. Besides, the observed discrepancy is also subject to some numerical influence due to differences among stellar evolution codes.

So far, the comparisons have been carried out primarily using Galactic GCs, hence exploring sub-solar metallicities and old ages only.
The possible analysis of the bump with seismic data has been suspected for a while \citep[see, e.g.,][]{Kallinger2010}. Thanks to asteroseismic constraints coupled with spectroscopic constraints, we are able to lead a distance-independent study of the RGBb using thousands of field stars, hence exploring a much larger domain of mass, age, and metallicity.


\begin{figure*}[ht!]
\resizebox{\hsize}{!}
{\plotone{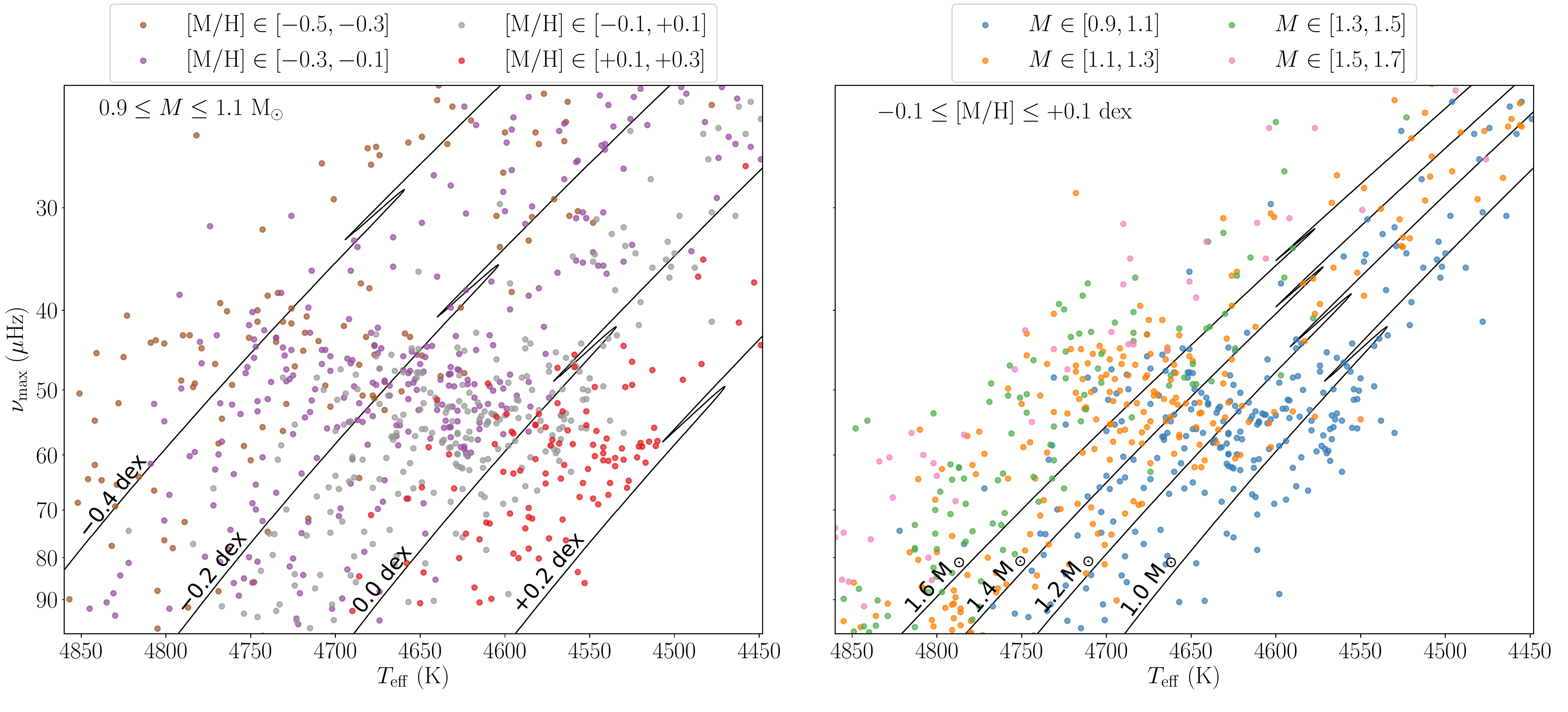}}
\caption{Left: $T_{\rm eff} - \nu_{\rm max}$ diagram, zoomed near the location of the bump. The coloured points correspond to the data (see Sec. \ref{sec:data}) with a mass $M \in \rm [0.9, 1.1] \ M_{\rm \odot}$, and different metallicity ranges: $\rm [-0.5, -0.3]$ (brown), $\rm [-0.3, -0.1]$ (purple), $\rm[-0.1, +0.1]$ (grey), $\rm [+0.1, +0.3] \ dex$ (red). Evolutionary tracks computed using MESA (see Sec. \ref{sec:models}) with $M = 1.0 \rm \ M_{\rm \odot}$, and $\rm{[M/H]} = -0.4, -0.2, 0.0, +0.2 \rm \ dex$ (from left to right), without envelope overshooting, are shown in black. Right: same diagram with a metallicity $\rm{[M/H]} \in [-0.1, +0.1] \ dex$, and different mass ranges: $\rm [0.9, 1.1]$ (blue), $\rm [1.1, 1.3]$ (orange), $\rm[1.3, 1.5]$ (green), $\rm [1.5, 1.7] \ M_{\rm \odot}$ (pink). Evolutionary tracks with $\rm{[M/H]} = 0.0 \rm \ dex$, and $M = 1.0, 1.2, 1.4, 1.6 \rm \ M_{\rm \odot}$ (from right to left), without envelope overshooting, are shown in black. \label{fig:teff_numax}}
\end{figure*}

\section{Observational and theoretical framework}

   \subsection{Data}
   \label{sec:data}

Our sample consists of red-giant stars observed by \textit{Kepler} and with APOGEE spectra \citep{SDSSCollaboration2016} available (APOKASC collaboration). From the initial list of stars, we select those that are classified as RGB using the method by \citet{Elsworth2017}.

We use the global asteroseismic parameters extracted from the frequency-power spectrum of the light curves by means of \citet{Mosser2011}'s data analysis method. These global seismic quantities are: the frequency of maximum oscillation power $\nu_{\rm max}$ and the average large frequency spacing $\langle \Delta \nu \rangle$. We also make use of the spectroscopically measured effective temperature $T_{\rm eff}$, with the required post-calibration (details available online \footnote{\url{http://www.sdss.org/dr13/irspec/parameters/}}), and of constraints on the photospheric chemical composition $\rm[Fe/H]$ and $\rm[\alpha/Fe]$ from SDSS DR13 \citep{SDSSCollaboration2016}. 
Furthermore, for stars showing enhancement in the $\alpha$ elements, we compute a metallicity [M/H] using the prescription described by \citet{Salaris1993} (which allows comparing to models calculated with solar-scaled abundances). Our final sample contains $\approx 3000$ RGB stars.

Stellar masses are inferred using the Bayesian tool PARAM 
\citep{Rodrigues2017}. Asteroseismic constraints $\nu_{\rm max}$ and \deltanu\ are included in the modelling procedure in a self-consistent manner, whereby \deltanu\ is calculated from a linear fitting of the individual radial-mode frequencies of the models in the grid. 
At this time, this approach has yielded masses / radii which show no systematic deviations to within a few percent of independent estimates (see, e.g., \citealt{Miglio2016}; \citealt{Rodrigues2017}; \citealt{Handberg2017}; \citealt{Brogaard2018}, who partially revisited the work by \citealt{Gaulme2016}). 

The effects of potential systematic biases in the mass and metallicity scale
are discussed in Sec. \ref{sec:systematics}. In our dataset, the typical random uncertainties are of the order of 1.95\% on $\nu_{\rm max}$, 0.05 $\mu$Hz on \deltanu, 70 K on $T_{\rm eff}$, 0.04-0.07 dex on $\rm[M/H]$, and 6-10\% on mass.

   \subsection{Models}
   \label{sec:models}

Evolutionary tracks and interior structures of RGB stars are computed using the stellar evolution code MESA \citep{Paxton2011, Paxton2013, Paxton2015}. 
We compute evolutionary tracks with $M$ ranging from $1.0$ to $1.6 \ \rm M_{\rm \odot}$ in steps of $0.2 \rm \ M_{\rm \odot}$, and $\rm[M/H]$ spanning from $-0.4$ to $0.2 \ \rm dex$ in steps of $0.2 \rm \ dex$. 
The initial helium mass fraction $Y_{\rm 0}$ is determined assuming a linear chemical enrichment law $\Delta Y / \Delta Z$, with $Z_{\rm \odot}=0.01756$ and $Y_{\rm 0, \odot}=0.26627$.
The mixing-length parameter is taken equal to the solar-calibrated value $\alpha_{\rm MLT} = 1.9658$. For more details about the physical inputs of the models, we refer the reader to \citet{Rodrigues2017}, with the exception that we include diffusive convective core overshooting $\alpha_{\rm ov,core}=0.01$ during the main sequence. In this work, we focus on a diffusive type of mixing \citep{Herwig2000}, with three different overshooting efficiencies below the lower boundary of the convective envelope: $\alpha_{\rm ov,env}=0.00$ (no overshooting), $\alpha_{\rm ov,env}=0.025$, and $\alpha_{\rm ov,env}=0.05$. These would correspond to models with a fully-mixed overshooting region of the order of $\sim 0.3 \rm \ H_{\rm P}$ for $\alpha_{\rm ov,env}=0.025$, and $\sim 0.6 \rm \ H_{\rm P}$ for $\alpha_{\rm ov,env}=0.05$.

Systematic effects due to different assumptions on the initial helium mass fraction, on the mixing-length parameter, and on convective core overshooting during the main sequence are presented in Sec. \ref{sec:systematics}.

Lastly, $\nu_{\rm max}$ is estimated through the seismic scaling relations, with the following solar references: $\nu_{\rm max, \odot} = 3090 \rm \ \mu Hz$ and $T_{\rm eff, \odot} = 5777 \rm \ K$. As for $\langle\Delta \nu\rangle$, we follow the average large frequency definition described by \citet{Rodrigues2017} (see Sec. \ref{sec:data}), where the individual radial-mode frequencies are computed with \texttt{GYRE} \citep{Townsend2013}.

The $T_{\rm eff} - \nu_{\rm max}$ diagrams of the data used in this work, overlaid with a few evolutionary tracks, are displayed in Fig. \ref{fig:teff_numax}, where the RGBb appears as a clear feature in $\nu_{\rm max}$.

It is worth noticing that the \textit{Kepler} and APOGEE target selection was primarily based on colour and magnitude criteria that, for the masses / metallicities explored in this study, are not expected to affect the recovered position of the RGBb \citep[see, e.g.,][]{Farmer2013, Miglio2014, Pinsonneault2014}. 
Moreover, we do not expect significant selection effects due to the length and cadence of \textit{Kepler}'s observations, in the $\nu_{\rm max}$ (hence $\log g$) range where we identify the bump. Biases against high-$\log g$ stars are expected if one were to extend the domain to frequencies closer to the Nyquist frequency of \textit{Kepler} long-cadence data ($\nu_{\rm max} \simeq 283 \rm \ \mu Hz$), which falls outside the domain relevant for our analysis. Also, if one were to extend to low-$\log g$ values, considerably lower than the red clump (e.g. $\nu_{\rm max} \lesssim 10 \rm \ \mu Hz$), biases may start to be significant due to the limited duration of the observations and to the target selection being biased against intrinsically luminous stars \citep[see, e.g.,][]{Farmer2013, Pinsonneault2014}.
 
It is also worth stressing that we have a sample that is little, if at all, contaminated by non-RGB stars, since core-helium burning stars have been identified and removed from the sample using the evolutionary-dependent signature of gravity modes in the oscillation spectra \citep{Bedding2011,Elsworth2017}.


\section{Determination of the RGB bump location}
\label{sec:rgbb_location_extent}

\begin{figure*}
\resizebox{\hsize}{!}
{\plottwo{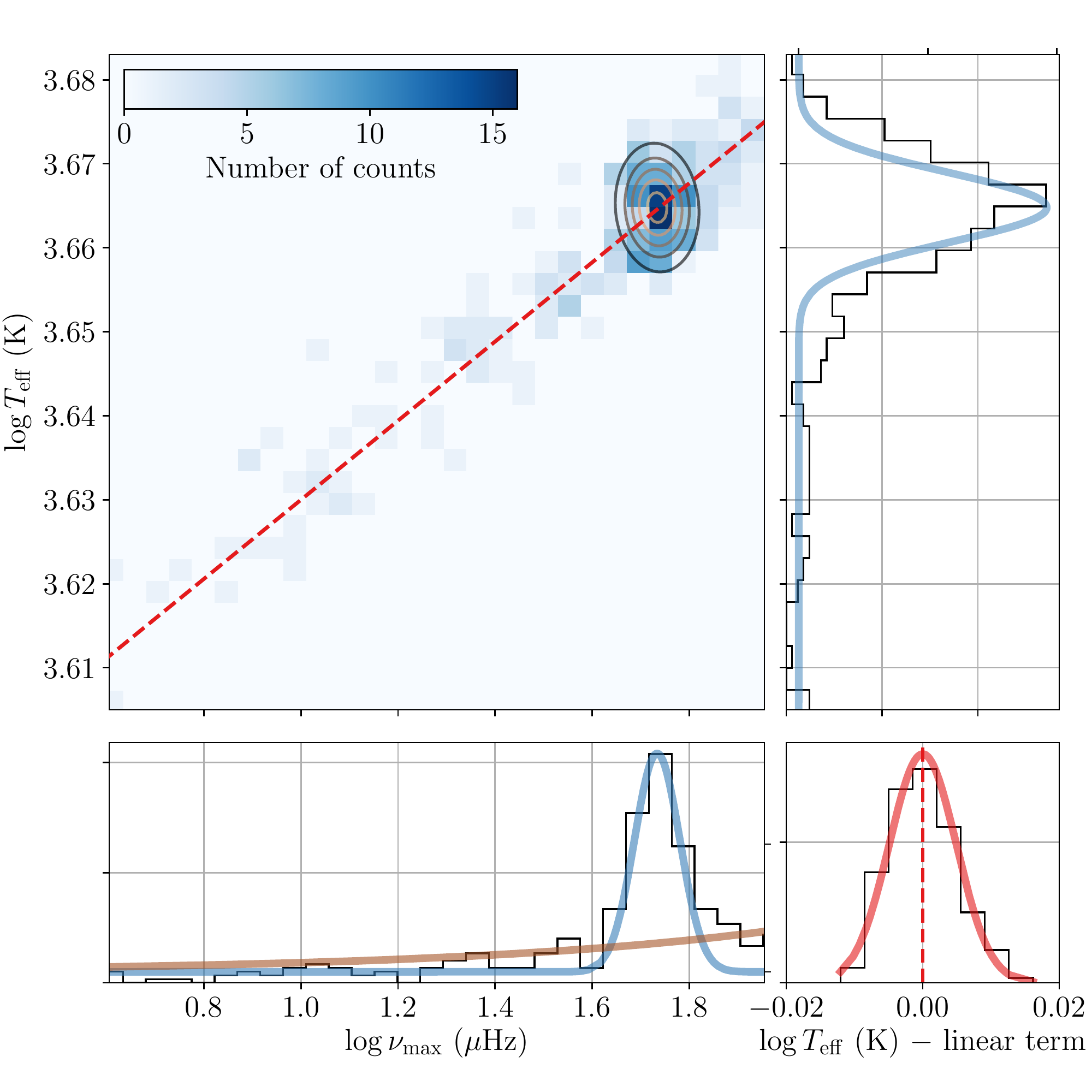}{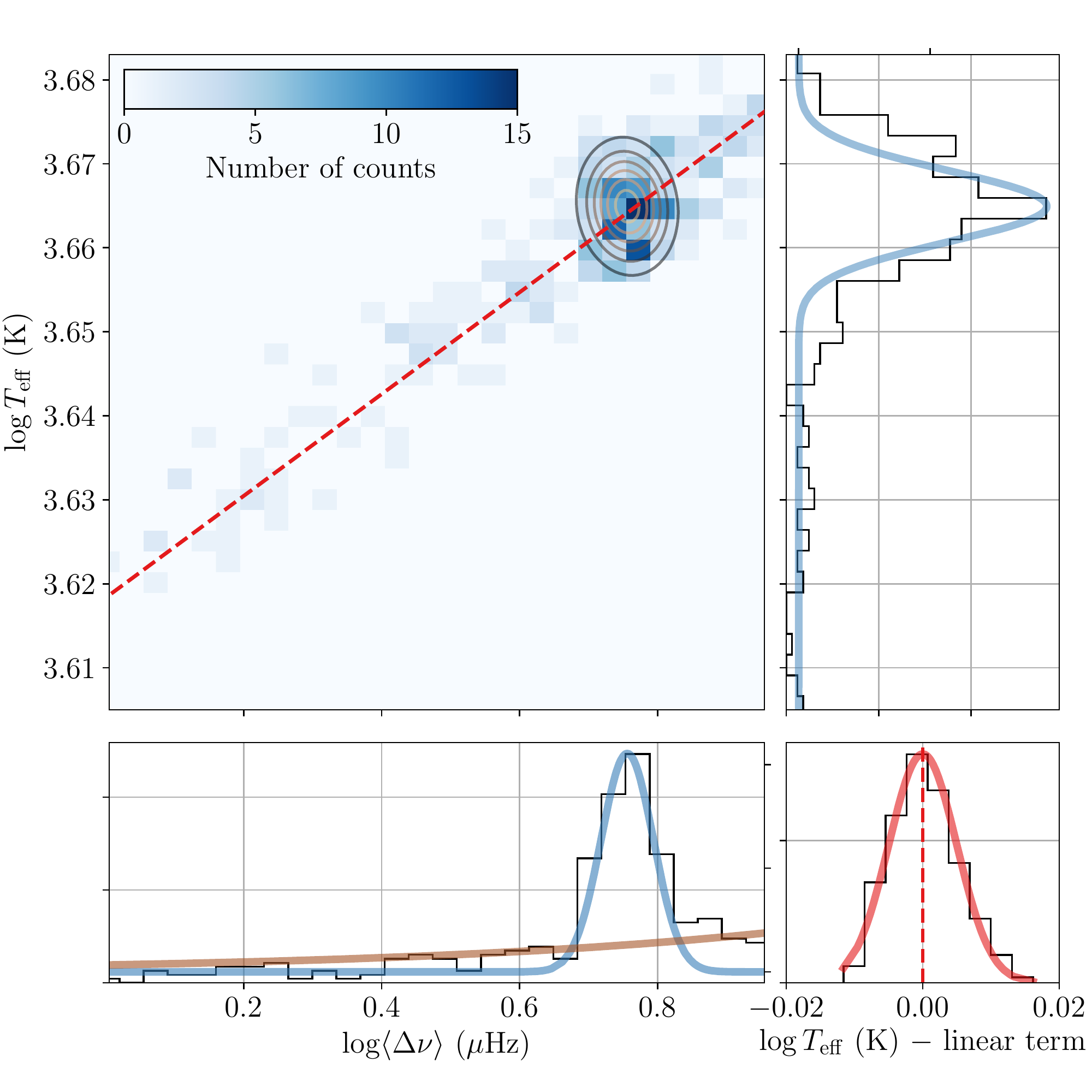}}
\caption{Probability functions applied to our dataset in the $\log T_{\rm eff} - \log \nu_{\rm max}$ (left) and $\log T_{\rm eff} - \log \deltanu$ (right) planes; with $0.9 \leq M \leq 1.1 \rm \ M_{\rm \odot}$ and $-0.1 \leq \rm{[M/H]} \leq +0.1 \rm \ dex$. 2D histograms are plotted, where the colour scale indicates the number of stars and the ellipses show the location of the bump, as determined by the mixture model technique. 1D histograms of $\log \nu_{\rm max}$ (or $\log \deltanu$) and $\log T_{\rm eff}$ are also shown in black in the bottom and right plots, respectively. The two components of the RGBb bivariate Gaussian are displayed with blue lines. The brown line correspond to the RGB outliers' rising exponential in $\log \nu_{\rm max}$ (or $\log \deltanu$); while the red dashed line shows the linear term modelling the RGB background. The small plot at the bottom right corner depicts the difference between $\log T_{\rm eff}$ and the RGB linear term, hereby illustrating the normal scatter. \label{fig:smm}}
\end{figure*}


To detect and characterise the RGBb, we use a statistical mixture model \citep{Hogg2010} to estimate its position 
in $\nu_{\rm max}$ and $\langle\Delta \nu\rangle$.   
    
The mixture model approach is a statistical framework allowing simultaneous consideration of multiple models, or hypotheses, with reference to a single dataset. In this context, they are a means of distinguishing inliers, the RGBb overdensity, and outliers, the RGB background, i.e. the remaining stars not belonging to the bump. We apply our fitting method in the $\log T_{\rm eff} - \log \nu_{\rm max}$ and $\log T_{\rm eff} - \log \langle\Delta \nu\rangle$ planes, where the RGBb appears as a dense and slightly sloping horizontal strip in a restricted bin of mass and metallicity.
    
We employ one common strategy for our dataset and for simple synthetic populations ---  mono-mass, mono-metallicity --- derived from MESA models. In both cases, three probability functions are at work: the RGBb foreground is described by a bivariate normal distribution with a negative correlation; while a rising exponential in $\log \nu_{\rm max}$ (or $\log \langle\Delta \nu\rangle$) and a linear term with a normally distributed scatter are used for the remaining population of RGB outliers. These probability functions are depicted in Fig. \ref{fig:smm}. Notably, considering $1/\nu_{\rm max}$ as a proxy of the luminosity ($1/\nu_{\rm max} \propto R^2$, where $R$ is the radius), one can appreciate that the behaviour of the RGB background follows the same trend as the luminosity function of GCs (with a decreasing number of stars with increasing luminosity).
    
The mixture model likelihood function is then marginalised using a Markov Chain Monte-Carlo process, by means of the Python package \texttt{emcee} \citep{Foreman-Mackey2013}, producing posterior probability distributions for our set of ten parameters: the RGBb location in $\log \nu_{\rm max}$ (or $\log \langle\Delta \nu\rangle$) and $\log T_{\rm eff}$, the corresponding RGBb standard deviations, the correlation of the bivariate Gaussian, the exponential parameter, the linear term's slope, intercept and standard deviation, and the mixture model weighting factor.


\section{Results}
\label{sec:results}

As the RGBb properties are expected to be highly dependent on stellar parameters, we divide our sample in the following intervals of mass: $[0.9, 1.1]$, $[1.1, 1.3]$, $[1.3, 1.5]$, and $[1.5, 1.7]$ $\rm M_{\rm \odot}$; and metallicity: $[-0.5, -0.3]$, $[-0.3, -0.1]$, $[-0.1, +0.1]$, and $[+0.1, +0.3]$ dex. Out of all 16 mass and metallicity bins, we can detect and robustly characterise the RGBb position in both $\nu_{\rm max}$ and $\langle\Delta \nu\rangle$ in nine of them, with the main limiting factor being the low number of stars in some of the bins. The typical uncertainties ($68\%$ credible region) on the RGBb position in $\nu_{\rm max}$ and \deltanu\ go from 1\% to 5\% for the most poorly populated bins.

   \subsection{Trends with mass and metallicity}

We investigate the location of the observed RGBb as a function of $M$ and $\rm [M/H]$. Both the $\nu_{\rm max}$ and $\langle\Delta \nu\rangle$ of the RGBb decrease with increasing stellar mass and decreasing metallicity (Fig. \ref{fig:undershoot_M} and \ref{fig:undershoot_Z}).

From theoretical models, one expects that hotter stars, as a result of a higher mass or a lower metallicity, have a shallower convective envelope, hence a higher RGBb luminosity. This increase in luminosity is typically accompanied by an increase in radius. The latter explains the decrease in $\nu_{\rm max}$ and $\deltanu$, which are strongly dependent on $R$ ($\nu_{\rm max} \propto R^{-2}$, $\deltanu \propto R^{-3/2}$).

In conclusion, whether in terms of $\nu_{\rm max}$ or $\langle\Delta \nu\rangle$, the data show a RGBb having trends with mass and metallicity that are qualitatively consistent with expectations. We now proceed to a quantitative comparison between observations and predictions.
   
   \subsection{A preliminary calibration of the envelope overshooting parameter}
   
\begin{figure*}[ht!]
\resizebox{\hsize}{!}
{\plotone{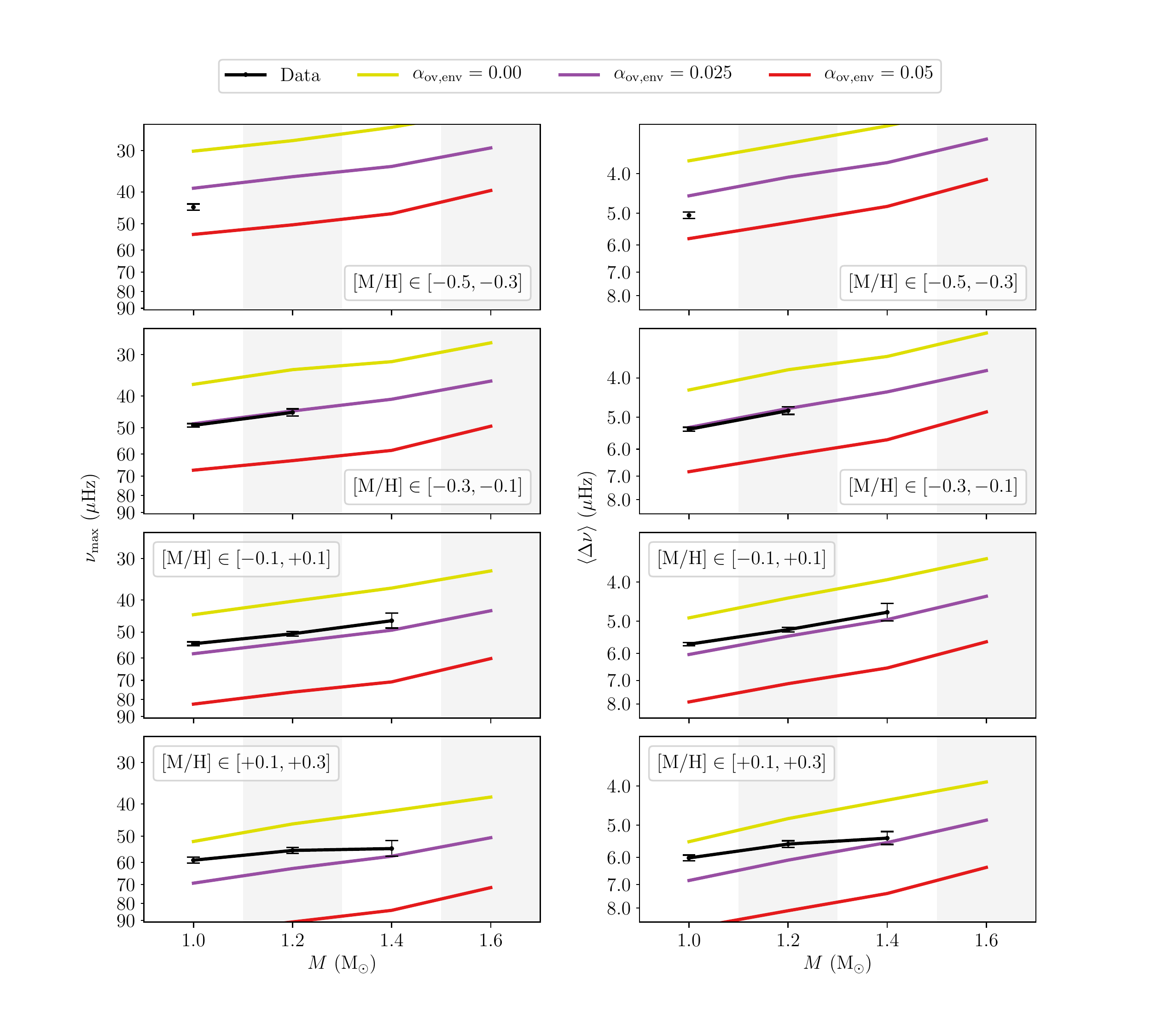}}
\caption{Location of the RGBb in $\nu_{\rm max}$ (left) and in $\langle\Delta \nu\rangle$ (right), in our dataset (black; see Sec. \ref{sec:data}), and its corresponding $68\%$ credible region, and simple synthetic populations with different envelope overshooting efficiencies: $\alpha_{\rm ov,env}=0.00$ (yellow), $\alpha_{\rm ov,env}=0.025$ (purple), and $\alpha_{\rm ov,env}=0.05$ (red); as a function of mass, for different metallicity ranges: $[-0.5, -0.3]$, $[-0.3, -0.1]$, $[-0.1, +0.1]$, and $[+0.1, +0.3] \rm \ dex$ (from top to bottom). The background bands indicate the mass bin, in which the bump position has been estimated. \label{fig:undershoot_M}}
\end{figure*}

\begin{figure*}[ht!]
\resizebox{\hsize}{!}
{\plotone{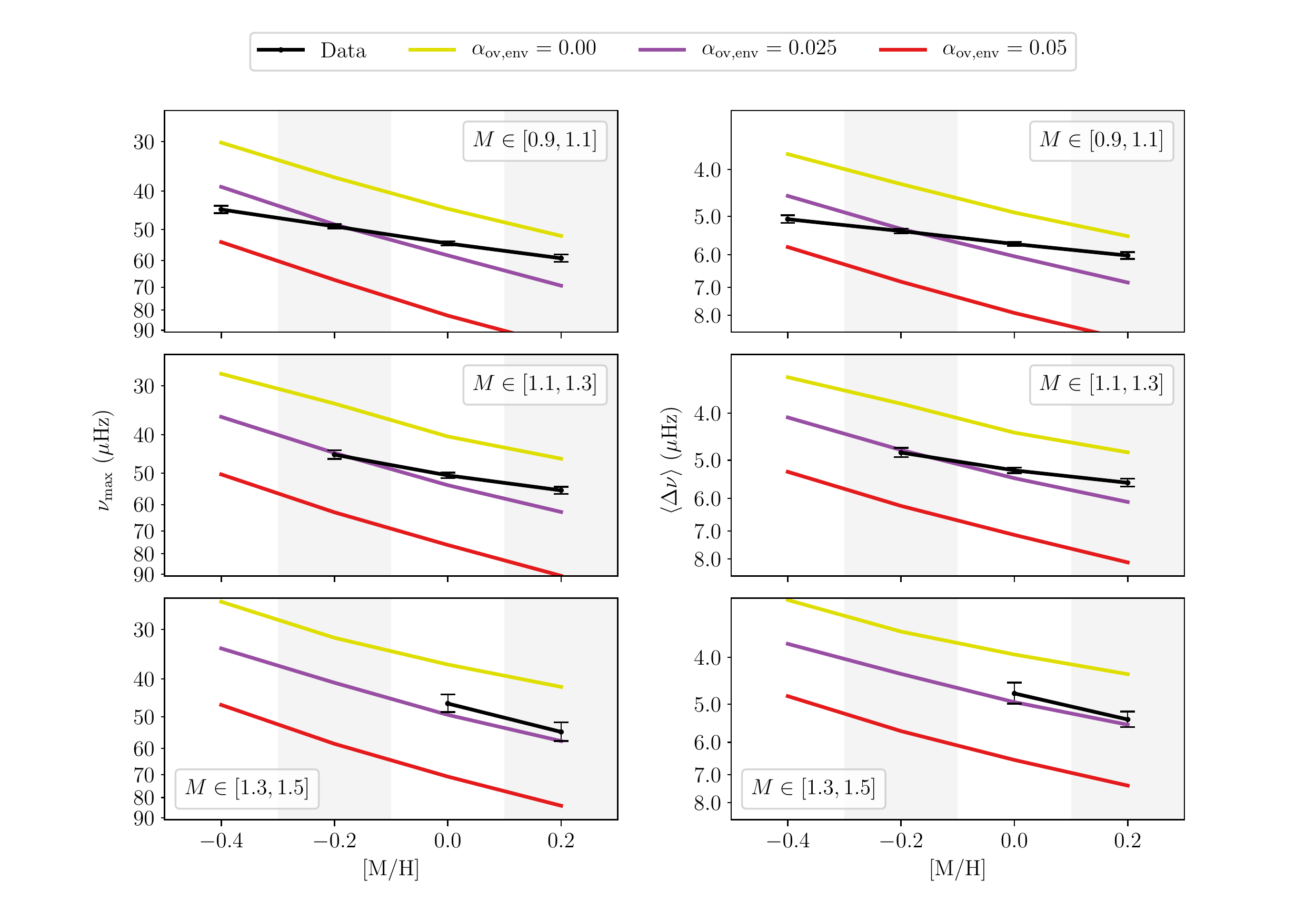}}
\caption{Same as Fig. \ref{fig:undershoot_M}; as a function of metallicity, for different mass ranges: $[0.9, 1.1]$, $[1.1, 1.3]$, and $[1.3, 1.5] \rm \ M_{\rm \odot}$ (from top to bottom). \label{fig:undershoot_Z}}
\end{figure*}

We generate simple synthetic populations --- mono-mass, mono-metallicity ---  from MESA models (see Sec. \ref{sec:models}). For each track, we create a population of 3000 stars, assuming a uniform age distribution, and we interpolate in age to get $\nu_{\rm max}$, $\langle\Delta \nu\rangle$, and $T_{\rm eff}$. A normally distributed noise is then added, using the typical uncertainties on each of these properties (see Sec. \ref{sec:data}).
   
We then compare the RGBb observed locations with those that are predicted by theory, for different envelope overshooting efficiencies. To take into account the latter, we use a diffusive type of mixing in MESA, characterised by an adjustable parameter $\alpha_{\rm ov,env}$ (see Sec. \ref{sec:models}).
   
The comparison of the observed and predicted 
RGBb mean value in $\nu_{\rm max}$ and $\langle\Delta \nu\rangle$, as a function of mass and metallicity, is displayed on Fig. \ref{fig:undershoot_M} and \ref{fig:undershoot_Z}. First and foremost, it is clear that one would need to consider models with significant overshooting from the base of the convective envelope to reproduce the observations for both global seismic parameters. This conclusion holds whether we tackle the issue in terms of mass (Fig. \ref{fig:undershoot_M}) or metallicity (Fig. \ref{fig:undershoot_Z}).
   
Subsequently, focusing our attention on $\nu_{\rm max}$, we note that the most metal-poor stars are likely to suggest a slightly more substantial overshooting efficiency, which would be close to $\alpha_{\rm ov,env}=0.025$ or even greater than that for stars with $\rm{[M/H]} \in [-0.5,-0.3]$. Besides, we also checked whether the trend at low metallicity is dominated by $\alpha$-rich stars, yet only considering stars with $\rm{[\alpha/Fe]} < 0.05 \rm \ dex$  had no influence on the observed trends. As we go towards higher metallicities, a shallower envelope overshooting ($\alpha_{\rm ov,env} < 0.025$) is suggested.
Similar conclusions are reached from $\langle\Delta \nu\rangle$. 
Our analysis clearly shows that, for the models considered here, the extra-mixing efficiency for low-mass stars ($M \leq 1.3 \rm \ M_{\rm \odot}$) decreases as metallicity increases.

Finally, it is also apparent that the suggested overshooting efficiency never goes as far as $\alpha_{\rm ov,env}=0.05$ --- equivalent to a fully-mixed overshooting region of $\sim 0.6 \rm \ H_{\rm P}$ --- within the limits of our dataset, implicitly defining an upper bound for the extent of mixing needed.

   \subsection{Assessing other systematic effects}
   \label{sec:systematics}

\begin{figure*}[ht!]
\resizebox{\hsize}{!}
{\plotone{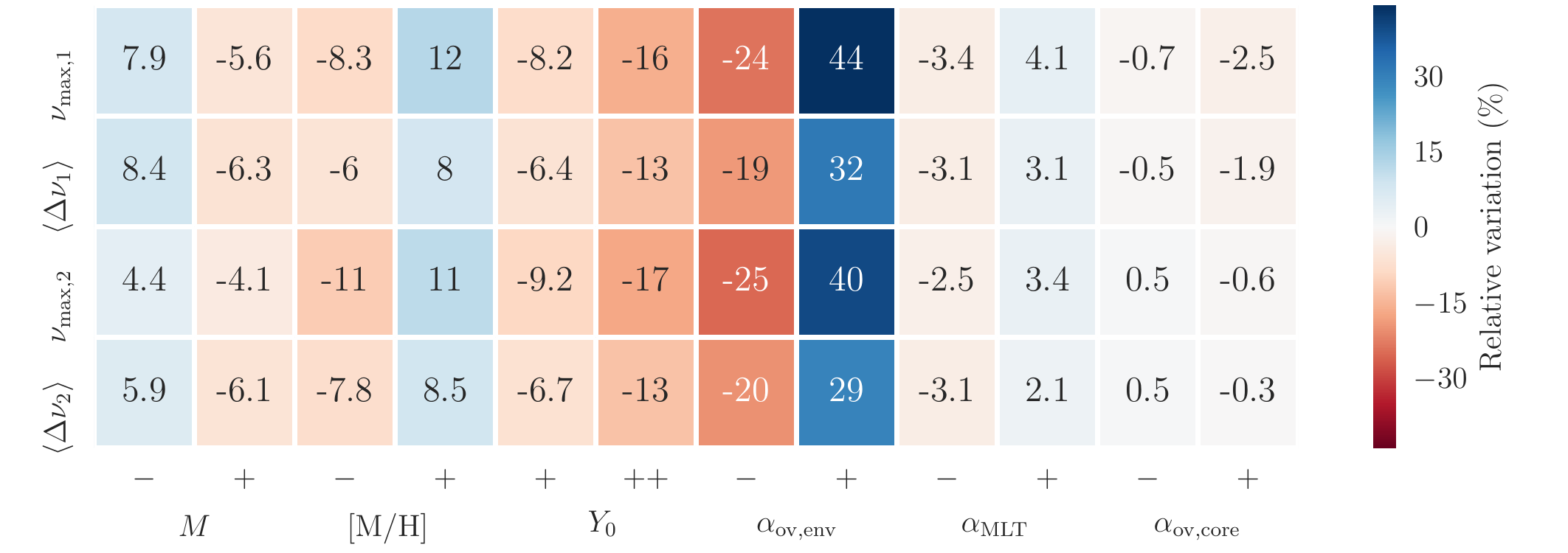}}
\caption{Heatmap illustrating the relative variations (\%) of the bump's location in $\nu_{\rm max}$ and $\langle\Delta \nu\rangle$, with two different sets of default parameters. Full details are provided in the text. \label{fig:heatmap}}
\end{figure*}

Uncertainties on additional parameters, other than the extent of mixing beyond the convective envelope, could also affect the position of the RGBb in $\nu_{\rm max}$ and $\langle\Delta \nu\rangle$. For this reason, we aim to assess the bias on the estimation of $\nu_{\rm max}$ and $\langle\Delta \nu\rangle$ induced by each of the following parameters: $M$, $\rm{[M/H]}$, $Y_{\rm 0}$, $\alpha_{\rm ov,env}$, $\alpha_{\rm MLT}$, and $\alpha_{\rm ov,core}$. 
To that end, all parameters are fixed excluding the one whose effect we wish to quantify. We test two sets of default parameters:
\begin{enumerate*}[label={\arabic*)},font={\color{black}\bfseries}]
\item $M=1.40 \rm \ M_{\rm \odot}$, $\rm{[M/H]} = 0.0 \rm \ dex$;
\item $M=1.20 \rm \ M_{\rm \odot}$, $\rm{[M/H]} = -0.2 \rm \ dex$.
\end{enumerate*}

A variation of $10 \%$ in mass affects the $\nu_{\rm max}$ and $\langle\Delta \nu\rangle$ of the bump in the order of $6 \%$. Changing the metallicity ($\pm 0.1 \rm \ dex$) or the initial helium mass fraction ($+0.02,0.04$) has more impact on $\nu_{\rm max}$ than on $\langle\Delta \nu\rangle$: $10 \%$ and $8.5 \%$ against $7 \%$ and $6 \%$, respectively. As expected, as $Y_{\rm 0}$ increases, stars are hotter and brighter \citep[e.g.][]{Fagotto1994}, and the RGBb ends up at a higher luminosity. Unquestionably, the envelope overshooting efficiency ($\pm0.025$) has the most significant effect on the position of the bump, again greater for $\nu_{\rm max}$: about $34 \%$ compared to $\sim 25 \%$ in $\langle\Delta \nu\rangle$. The mixing-length parameter ($\pm0.1$) has a mild repercussion, of a few percent, whether on $\nu_{\rm max}$ or $\langle\Delta \nu\rangle$. Moreover, both fits of colour-magnitude diagrams with GCs and hydrodynamics simulations do not suggest changes in $\alpha_{\rm MLT}$ significantly larger than $\pm 0.1$ \citep[see, e.g.,][]{Trampedach2014, Salaris2018}. Finally, the convective core overshooting efficiency ($\pm 0.01$) during the main-sequence phase does not have any notable effect on the RGBb position for a mass and a metallicity up to $1.4 \rm \ M_{\rm \odot}$ and $0.3 \rm \ dex$, which draw the limits of the current work's dataset. The relative variations are portrayed with a heatmap in Fig. \ref{fig:heatmap}.
   

A combination of each of these parameters might be able to account for the discrepancy between the observed and predicted RGBb position, but it seems unlikely. 

The helium-to-metals enrichment ratio (see Sec. \ref{sec:models}) may have a significant effect on the estimate of the extra mixing required to fit the observations. Indeed, at fixed 
$Z$, an increase in $\Delta Y / \Delta Z$ induces a greater initial helium abundance, thus a decrease in the $\nu_{\rm max}$ and $\deltanu$ of the bump (see Fig. \ref{fig:heatmap}). This $Y_{\rm 0}$ increase is further amplified at high metallicities, hence the corresponding conclusions might suggest a greater amount of overshooting from the lower boundary of the convective envelope. 
If one were to explore lower values of $Y_{\rm 0}$, the discrepancy would be reduced but these would have to be sub-big-bang-nucleosynthesis values \citep[see][]{Troisi2011}.


\section{Conclusions}

First of all, we report the detection of the RGBb in stars observed by {\it Kepler} with APOGEE spectra, corresponding to the widest mass and metallicity domain explored thus far. Previous studies of the bump in GGCs could hardly illustrate the expected trends of the RGBb luminosity with mass and metallicity, because of their analysis being restricted to predominantly clusters with sub-solar metallicities. The combination of asteroseismology and spectroscopy provides us with the resources required to finally overcome this barrier. 

We show that the observed RGBb location in $\nu_{\rm max}$ and $\langle\Delta \nu\rangle$ reveals trends with mass and metallicity in line with expectations from models. However, we note that models without envelope overshooting are in disagreement with observations. 
This result is confirmed whether we approach the problem in terms of $\nu_{\rm max}$ or $\langle\Delta \nu\rangle$. Indeed, the most metal-poor stars seemingly suggest a mixing extent of $\alpha_{\rm ov,env} \geq 0.025$; while a lower amount of overshooting, $0.00 \leq \alpha_{\rm ov,env} \leq 0.025$, seems more appropriate for their metallicity-enhanced counterparts. Hence, for both seismic observables, we see hints of a possible dependence of the extra-mixing efficiency on metallicity. Finally, for the sample considered in this work, a strong efficiency such as $\alpha_{\rm ov,env}=0.05$ can be set aside, showing that there is an upper limit to the extent of mixing needed to bring an agreement between models and observations.

Interestingly, similar evidence for additional mixing is found in GCs. \citet{Cassisi2011} and \citet{Fu2018} mentioned that the offset between predicted and observed RGBb would disappear with the inclusion of overshooting in the order of $0.25 \rm \ H_{\rm p}$ at the base of the convective envelope. These works, as well as \citet{DiCecco2010}, \citet{Troisi2011} and \citet{Joyce2015}, also found an increasing discrepancy when moving from metal-rich to metal-poor GCs. To alleviate this discrepancy, \citet{Fu2018} mentioned the genuine possibility for the extra-mixing efficiency to be larger than their adopted value in metal-poor stars. This assumption seems to tally with our preliminary finding of a possible dependence of envelope overshooting on metallicity.
   
Furthermore, our conclusion is strengthened by an assessment of systematic effects on the RGBb position in $\nu_{\rm max}$ and $\deltanu$, presented in Sec. \ref{sec:systematics}. That said, it is crucial to bear in mind that further tests are needed for a robust calibration of the amount of extra mixing from the base of the convective envelope. These tests should exhaustively consider all conceivable systematic biases, including the comparison among various stellar evolution codes and, e.g., different helium-to-metals enrichment rates $\Delta Y / \Delta Z$.

In this regard, the second Data Release of \textit{Gaia} will allow a significant step forward in characterising the bump. With precise and accurate parallaxes available, one can use additional diagnostics to avoid, or mitigate, theoretical and observational uncertainties, e.g. by comparing the RGBb luminosity with that of the zero-age horizontal branch or the main-sequence turn off.

Moreover, by considering the full \textit{Kepler} dataset, and data from K2 and TESS, it will soon be possible to couple astrometric and spectroscopic constraints with asteroseismic data in significantly larger samples of stars, thus providing further insights into the efficiency of internal mixing processes in cool stars.

\begin{acknowledgments}
AM, GRD, BM and LG are grateful to the International Space Science Institute (ISSI) for support provided to the \mbox{asteroSTEP} ISSI International Team.
OJH, AM and GRD acknowledge the support of the UK Science and Technology Facilities Council (STFC). LG and JM acknowledge support from the ERC Consolidator Grant funding scheme ({\em project STARKEY}, G.A. n. 615604). AM acknowledges support from the ERC Consolidator Grant funding scheme ({\em project ASTEROCHRONOMETRY}, G.A. n. 772293). We also wish to thank the referee, Achim Weiss, for his comments, that helped clarify and improve the paper.

\end{acknowledgments}

\bibliographystyle{aasjournal}

\end{document}